\documentclass[final,3p,times]{elsarticle}

\journal{Technical Report}

\usepackage{hyperref}
\makeatletter
\g@addto@macro{\UrlBreaks}{\UrlOrds}
\makeatother

\makeatletter \newif\if@restonecol \makeatother  
\usepackage{algorithm}
\usepackage{algorithmic}
\usepackage{eurosym}

\usepackage[dvipsnames]{xcolor}
\usepackage{tabularx}
\usepackage{soul}
\usepackage{amssymb}
\usepackage{pifont}

\usepackage{amsmath}
\usepackage{amsfonts}

\usepackage{bbm}
\usepackage{dsfont}
\usepackage{bm}
\usepackage{xspace}
\usepackage{paralist}

\usepackage{paralist}
\usepackage{booktabs} %
\usepackage{multirow}
\usepackage{adjustbox}
\usepackage{placeins}
\usepackage{subcaption}
\usepackage{epstopdf}
\usepackage{graphics}
\DeclareGraphicsExtensions{.eps,.gz,.eps,.pdf,.png,.jpg}
\graphicspath{{./}{./figures/}}

\usepackage{graphicx}

\newcommand{\todo}[1]{#1}
 \newcommand{\edit}[1]{#1}

\newcommand{\sstitle}[1]{\smallskip\noindent\textbf{#1.\/}}

\newcommand{\toolname}{\texttt{FORTRESS}\xspace}

\def\Snospace~{\S{}}

\makeatletter
\newcommand{\removelatexerror}{\let\@latex@error\@gobble}
\makeatother

\begin{document}

\setlength{\belowdisplayskip}{2pt}
\setlength{\belowdisplayshortskip}{2pt}
\setlength{\abovedisplayskip}{2pt}
\setlength{\abovedisplayshortskip}{2pt}

\begin{frontmatter}

\title{Robust Aggregation for Federated Sequential Recommendation with Sparse and Poisoned Data}

\author{Minh Hieu Nguyen et al.}

\begin{abstract}
Federated sequential recommendation distributes model training across user devices so that behavioural data remains local, reducing privacy risks. Yet, this setting introduces two intertwined difficulties. On the one hand, individual clients typically contribute only short and highly sparse interaction sequences, limiting the reliability of learned user representations. On the other hand, the federated optimisation process is vulnerable to malicious or corrupted client updates, where poisoned gradients can significantly distort the global model. These challenges are particularly severe in sequential recommendation, where temporal dynamics further complicate signal aggregation.
To address this problem, we propose a robust aggregation framework tailored for federated sequential recommendation under sparse and adversarial conditions. Instead of relying on standard averaging, our method introduces a defence-aware aggregation mechanism that identifies and down-weights unreliable client updates while preserving informative signals from sparse but benign participants. The framework incorporates representation-level constraints to stabilise user and item embeddings, preventing poisoned or anomalous contributions from dominating the global parameter space. In addition, we integrate sequence-aware regularisation to maintain temporal coherence in user modelling despite limited local observations.
\end{abstract}


\end{frontmatter}

\section{Introduction}

Sequential recommender systems (SeqRecs) are widely used in real-world applications, including e-commerce, news platforms, and streaming services, where modeling users' temporal preferences is essential for delivering personalized and timely recommendations~\cite{rw_15,rw_17}. Traditional SeqRecs are usually trained in a centralized fashion that requires full access to user--item interaction logs. Although effective, this setting raises serious privacy concerns and is increasingly incompatible with modern data protection regulations. Federated learning has emerged as a promising solution by enabling collaborative model training without sharing raw user data~\cite{rw_22}. Building on this, recent studies in Federated Sequential Recommendation (FedSeqRec)~\cite{fmss, ptf-fsr} have demonstrated the potential of preserving user privacy while modeling temporal behaviors. We refer to the concrete model implementations operating in this setting as Federated Sequential Recommendation Systems (FedSRS).

\edit{
Despite recent progress, the challenges of data sparsity and model poisoning are particularly pronounced in federated sequential recommendation due to the combined effects of decentralization and temporal dependency. 
Unlike centralized sequential recommendation, where long interaction histories and global statistics can be aggregated across users, federated learning restricts each client to only its local interaction sequence, which is often short, fragmented, and highly sparse. 
This substantially weakens the supervision signal for learning reliable sequential representations, especially for cold-start or low-activity users.
}

In principle, contrastive learning can enrich supervision for sparse and noisy data and has shown strong gains in centralized recommendation~\cite{rw_49}. 
\edit{
Recent sequential recommendation methods further exploit multimodal pre-training and multi-view contrastive learning to alleviate sparsity and improve representation quality~\cite{zhang2024multimodal,xu2025multi}. 
However, these techniques typically rely on centralized access to cross-user signals (e.g., global negatives or shared interaction graphs), which cannot be directly applied in federated learning without violating privacy constraints or incurring significant communication overhead.
}
\edit{
Consequently, contrastive learning in existing FedSRS is either weakly instantiated as a local regularizer or entirely absent ~\cite{rw_37}.
Recent federated contrastive frameworks such as FedCSR~\cite{zheng2025fedcsr} focus on aligning local and global representations across platforms to handle cross-domain heterogeneity, but they assume benign clients with richer local data and do not address user-level sparsity or adversarial manipulation, limiting their applicability to realistic FedSRS settings.
}

\edit{
Moreover, federated sequential recommendation is inherently sensitive to temporal dynamics, where user preferences may change abruptly while each client only observes limited local history. 
The lack of global temporal context causes FedSRS models to overfit recent noise or overlook meaningful intent shifts, and simultaneously exposes them to targeted model poisoning attacks. 
Recent studies~\cite{a-hum,psmu} show that attackers can exploit temporal correlations by injecting short trigger patterns or manipulating popularity trajectories to bias next-item predictions while remaining statistically inconspicuous. 
Existing defenses mainly rely on aggregation-based filtering strategies originally developed for static recommendation, where interactions are treated independently and temporal order is ignored~\cite{rw_31,rw_32}. 
More broadly, although federated learning research has proposed mechanisms to mitigate client heterogeneity and feature shift (e.g., prototype learning and representation or parameter-space alignment)~\cite{fu2025federated}, these methods do not model how user or item representations evolve over time, nor do they explicitly consider temporally coordinated poisoning attacks. 
As a result, they are ineffective against promotion and camouflage attacks that gradually manipulate sequential patterns in federated sequential recommendation.
}

To overcome these limitations, we propose \toolname -- a novel \textbf{F}ederated c\textbf{O}ntrastive \textbf{R}obus\textbf{T} \textbf{RE}commender for \textbf{S}equential \textbf{S}ystems. \toolname is a communication-efficient, privacy-preserving, and adversarially robust framework designed to enhance both recommendation effectiveness and security in the FedSeqRec setting. On each client, \toolname generates simple augmented versions of the local interaction sequences and optimizes a multi-view contrastive objective over user, item, and sequence perspectives. These augmented sequences expose the model to more plausible behaviour patterns and help it learn more reliable representations even when user histories are short or sparse. To avoid overreacting to noise, \toolname further introduces a temporal consistency regularizer that discourages abrupt changes in the learned preferences across training rounds, allowing the model to quickly capture short-term interests while still preserving stable long-term tastes. Because this additional flexibility can also make the system easier to attack, \toolname complements local training with a lightweight server-side defense that monitors popularity shifts and penalizes suspicious promotion surges. We conduct extensive experiments on three real-world datasets -- Amazon Cell Phone, Amazon Baby, and MIND -- under both benign and adversarial settings. Empirical results show that \toolname consistently outperforms state-of-the-art baselines on these benchmarks in recommendation accuracy and robustness. Ablation and sensitivity studies further validate the complementary benefits and stability of each component.

In summary, this paper makes the following contributions:
\begin{itemize}
    \item We propose \toolname, a novel FedSRS framework that integrates multi-view contrastive learning with temporal and popularity-aware regularization to improve both personalization quality and adversarial resilience.
    \item We design a privacy-compliant contrastive learning strategy for FedSeqRec that enriches supervision for short and sparse user histories.
    \item We introduce a lightweight server-side defense mechanism that mitigates promotion attacks by modeling popularity-aware regularization over client updates.
    \item We conduct comprehensive evaluations on three datasets under adversarial conditions, demonstrating the effectiveness, robustness, and practical viability of \toolname.
\end{itemize}

Further details can be found in~\cite{nguyen2026handling}.

\section{Model and Problem Formulation}
\label{sec:preliminaries}

\sstitle{Problem Statement}
We consider a federated sequential recommendation, where a set of users $U = \{u_1, u_2, ..., u_N\}$ interact with a global item catalog $V = \{v_1, v_2, ..., v_M\}$. Each user $u \in U$ has a private interaction sequence $S_u = [v_1^u, v_2^u, ..., v_T^u]$ ordered by time. The goal is to train a model $f_\theta$ that predicts the next item $\hat{v}_{T+1}^u$ likely to be consumed by user $u$:
\[
\hat{v}_{T+1}^u = f_\theta(S_u).
\]
In the federated setting, raw interaction sequences stay on user devices; only model parameters are exchanged with the server during training.

\sstitle{Federated Optimization Protocol}
We adopt a synchronous federated learning process with repeated communication rounds between the server and sampled clients. At round $t$, the server broadcasts the global model $\theta^{(t)}$ to a subset of users $\mathcal{U}_t$. Each user $u \in \mathcal{U}_t$ runs $E$ local epochs to minimize
\[
\mathcal{L}_u = \mathcal{L}^{\text{Rec}}_u + \lambda_{\text{CL}} \mathcal{L}^{\text{CL}}_u + \lambda_{\text{TCR}} \mathcal{L}^{\text{TCR}}_u,
\]
where $\mathcal{L}^{\text{Rec}}_u$ is the next-item prediction loss, $\mathcal{L}^{\text{CL}}_u$ is the multi-view contrastive loss, and $\mathcal{L}^{\text{TCR}}_u$ is the temporal regularization.

After local training, client $u$ sends its updated parameters $\theta_u^{(t)}$ to the server. The server aggregates local models using a standard FedAvg operator:
\[
\theta^{(t+1)} = \text{Aggregate}\left(\{\theta_u^{(t)}\}_{u \in \mathcal{U}_t}\right)
= \sum_{u \in \mathcal{U}_t} \frac{n_u}{\sum_{k \in \mathcal{U}_t} n_k} \, \theta_u^{(t)},
\]
where $n_u$ is the number of interactions of user $u$ (or the local sample size).

After aggregation, the server applies a regularization objective on item embeddings:
\[
\mathcal{L}^{\text{server}} = \lambda_{\text{sep}} \mathcal{L}^{\text{sep}} + \lambda_{\text{var}} \mathcal{L}^{\text{var}},
\]
which combines global separation and variance stabilization to mitigate promotion and camouflage attacks at the global model level. 
\edit{To improve clarity and facilitate readability, we summarize all frequently used symbols and notations in \autoref{tab:notation-summary}.}

\begin{table}[t]
\centering
\caption{\edit{Summary of notations used.}}
\label{tab:notation-summary}
\footnotesize
\begin{tabular}{cl}
\toprule
Notation & Description \\
\midrule
$U, V$ & Sets of users and items \\
$S_u$ & Local interaction sequence of user $u$ \\
$f_\theta$ & Sequential recommendation model \\
$\theta^{(t)}, \theta_u^{(t)}$ & Global and local parameters at round $t$ \\
$n_u$ & Number of interactions of user $u$ \\
$\mathcal{L}^{\text{Rec}}_u$ & Next-item prediction loss \\
$\mathcal{L}^{\text{CL}}_u$ & Multi-view contrastive loss \\
$\mathcal{L}^{\text{TCR}}_u$ & Temporal regularization loss \\
$\mathcal{L}_u$ & Total client-side objective \\
$\mathcal{V}_{\text{hot}}, \mathcal{V}_{\text{sp}}$ & Popular and suspicious item sets \\
$\mathcal{L}^{\text{sep}}, \mathcal{L}^{\text{var}}$ & Server separation and variance losses \\
$\mathcal{L}^{\text{server}}$ & Total server-side regularization loss \\
\bottomrule
\end{tabular}
\end{table}

\sstitle{Model Poisoning Attacks in Federated Sequential Recommendation}
Although federated sequential recommendation keeps interaction sequences on local devices, malicious clients can still upload poisoned model updates that bias the global model toward specific target items. These model poisoning attacks are especially harmful in sequential recommendation, where next-item prediction depends on temporal patterns that can be manipulated while remaining hard to detect.

We focus on targeted model poisoning attacks, where attackers aim to promote a set of target items $\mathcal{V}^{\star}$ into the top-$K$ recommendations of many users. In sequential settings, this can be done by (i) injecting poisoned updates that create false next-item patterns (whenever a short trigger sequence appears, the model predicts the target item), or (ii) gradually pulling the target into popular clusters so it gains exposure through similar-item recommendations.

We measure attack success by the exposure ratio at rank $K$ (ER@$K$), the proportion of eligible users whose top-$K$ list contains the target item:
\[
\text{ER@}K = \frac{1}{|\mathcal{V}^{\star}|} 
\sum_{v_j \in \mathcal{V}^{\star}} 
\frac{|\{u_i \in U \mid v_j \in \hat{\mathcal{V}}_i,\; v_j \in \mathcal{V}_i^{-}\}|}
     {|\{u_i \in U \mid v_j \in \mathcal{V}_i^{-}\}|},
\]
where $\hat{\mathcal{V}}_i$ is the top-$K$ recommendation list of user $u_i$ and $\mathcal{V}_i^{-}$ is the set of items not yet consumed by $u_i$.

To construct strong and realistic attack scenarios, we adopt two recent methods. A-hum constructs hard users from Gaussian noise and generates poisoned updates that drag the target item toward popular items, acting as a camouflage attack. PSMU synthesizes pseudo-users with random but plausible sequences and adds alternative items to weaken competitors, acting as a promotion attack. Both methods require only a small number of malicious clients and no prior knowledge of the global dataset, making them practical and challenging baselines for testing the robustness of FedSRS~\cite{pham2024dual,nguyen2024multi,nguyen2024handling,yang2024pdc,sakong2024higher,huynh2024fast,huynh2025certified}.

\section{Methodology}
\label{sec:methodology}

\begin{figure*}[t]
    \centering
    \includegraphics[width=1.0\linewidth]{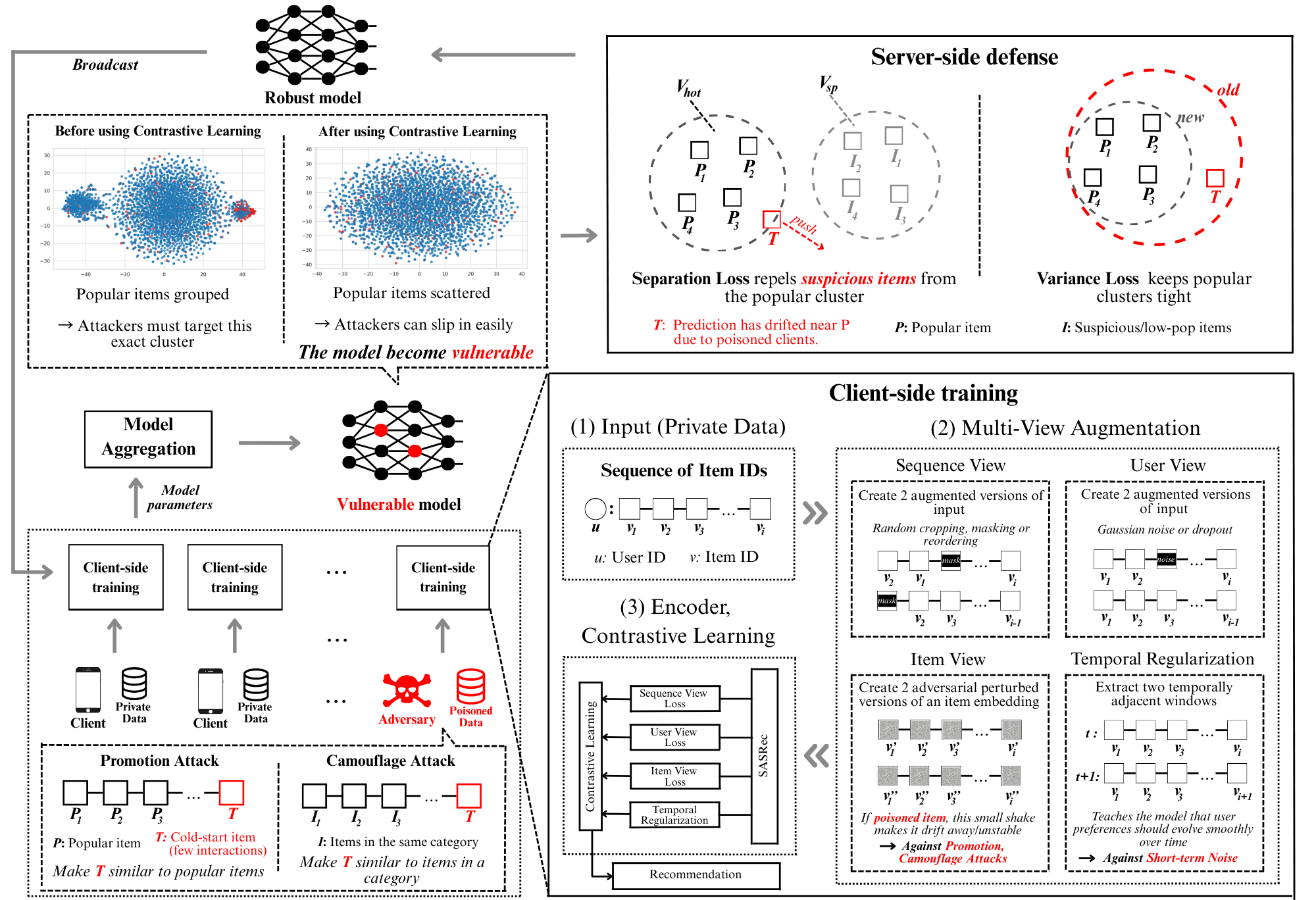}
    \caption{Overall architecture of the proposed FORTRESS framework.}
    \label{fig:framework}
\end{figure*}

In this section, we present \toolname, a federated learning framework tailored for sequential recommendation under privacy constraints. As shown in \autoref{fig:framework}, \toolname jointly addresses three challenges: (i) sparse and noisy local data, (ii) temporal instability in user preferences, and (iii) vulnerability to adversarial attacks, via three components: (1) contrastive learning for robust representations, (2) temporal regularization for preference stability, and (3) server-side defense for adversarial resilience. We first introduce the multi-view contrastive learning paradigm, then describe the temporal consistency regularization and server-side defense, and finally summarize the end-to-end training process in \autoref{alg:fortress}.

\subsection{Multi-View Contrastive Learning}

A core component of \toolname is \emph{multi-view contrastive learning} in the federated sequential recommendation setting. To obtain robust representations from heterogeneous and noisy decentralized data, we align multiple augmented views of the same underlying user preferences. We construct three contrastive views: a \emph{sequence view}, a \emph{user view}, and an \emph{item view}.
\todo{
In contrast to existing federated contrastive learning approaches, the contrastive design in \toolname{} differs in both supervision locality and augmentation strategy. While FedCL~\cite{rw_51} and UNION~\cite{rw_37} construct contrastive objectives by matching local representations against global references, and PTF-FSR~\cite{ptf-fsr} performs contrastive denoising over server-aggregated prediction sequences,
\toolname{} performs contrastive learning \emph{entirely on-device}. All contrastive views are generated locally from raw interaction sequences and embeddings, and no representations, contrastive pairs, or prediction traces are shared with the server~\cite{nguyen2023isomorphic,nguyen2024portable,ren2024comprehensive}.
}

\sstitle{Sequence View}
\todo{
Unlike prior methods that rely on global negatives, shared interaction graphs, or server-side sequence aggregation, the sequence view in \toolname{} uses only lightweight stochastic augmentations of local interaction histories. This design enables effective contrastive supervision under strict privacy constraints, while avoiding the exposure of sequential patterns that could be exploited by targeted poisoning attacks.
}
The sequence view focuses on noise and inconsistencies in user behavior sequences, which arise from fragmented histories, missing clicks, or UI-induced reordering. We enforce representation consistency across perturbed versions of the same interaction history. Specifically, we generate two augmented sequences by applying stochastic transformations to the original sequence $S_u$:
\[
\tilde{S}_u^{(1)} = \text{Aug}(S_u), \quad \tilde{S}_u^{(2)} = \text{Aug}(S_u).
\]
Each augmented sequence is then passed through the sequential encoder $f_\theta$ (e.g., SASRec or GRU4Rec), producing latent representations:
\[
\mathbf{h}_u^{(1)} = f_\theta(\tilde{S}_u^{(1)}), \quad \mathbf{h}_u^{(2)} = f_\theta(\tilde{S}_u^{(2)}).
\]
We optimize a sequence-level contrastive loss based on the InfoNCE formulation:
\[
\mathcal{L}^{\text{sc}}_u = \text{InfoNCE}(\mathbf{h}_u^{(1)}, \mathbf{h}_u^{(2)}).
\]
This encourages the encoder to focus on the core intent behind a sequence and be less sensitive to small perturbations and incomplete logs.

\sstitle{User View}
The user view addresses the gap between short-term behaviors and long-term preferences, often referred to as session drift. In federated settings, the server only sees local, temporally limited segments of each user, so it is important to stabilize identity-level embeddings. For a given user $u$, we first encode the full interaction sequence $S_u$ using the local model $f_\theta$, obtaining a latent user representation $\mathbf{u}_i = f_\theta(S_u)$. We then generate two perturbed versions of this embedding via stochastic noise:
\[
\mathbf{u}_i', \mathbf{u}_i'' = \text{Noise}(\mathbf{u}_i).
\]
We define a contrastive loss based on InfoNCE to align these noisy realizations:
\[
\mathcal{L}^{\text{uc}}_u = -\log \frac{\exp(\text{sim}(\mathbf{u}_i', \mathbf{u}_i'')/\tau)}{\sum_{j} \exp(\text{sim}(\mathbf{u}_i', \mathbf{u}_j)/\tau)}.
\]
By aligning different perturbations of the same user embedding, the user view promotes session-invariant user representations that are less affected by transient deviations.

\sstitle{Item View}
The item view aims to improve item representations under cold-start sparsity and sequential-aware poisoning. Cold-start items have few interactions, especially on each client, leading to unstable embeddings. At the same time, sequential poisoning attacks such as \textsc{A-hum}~\cite{a-hum} and \textsc{PSMU}~\cite{psmu} can exploit this instability by inserting target items into crafted sequences to bias next-item predictions. 
\todo{
Different from PTF-FSR\cite{ptf-fsr}, which applies contrastive learning after aggregation to smooth shared prediction sequences, the item view in \toolname{} directly regularizes item embeddings during local training. This client-side contrastive regularization explicitly stabilizes representation evolution and reduces the impact of poisoned gradients before model aggregation.
}
To stabilize item embeddings and reduce the impact of adversarial updates, we apply contrastive regularization to item representations. For each item $v_k$ in a user's sequence $S_u$, we simulate small perturbations using the gradient of the recommendation loss:
\[
\mathbf{v}_k', \mathbf{v}_k'' = \mathbf{v}_k + \nabla_{\mathbf{v}_k} \mathcal{L}^{\text{Rec}}_u.
\]
We then align the perturbed items with a contrastive loss:
\[
\mathcal{L}^{\text{ic}}_u = \sum_{v_k \in S_u} -\log \frac{\exp(\text{sim}(\mathbf{v}_k', \mathbf{v}_k'')/\tau)}{\sum_{v_j \in S_u} \exp(\text{sim}(\mathbf{v}_k', \mathbf{v}_j'')/\tau)}.
\]
This item-level objective encourages local consistency in the embedding space, smoothing the evolution of cold-start items and limiting embedding drift caused by poisoned gradients.

\subsection{Temporal Regularization}

User preferences in sequential recommendation usually change gradually rather than jumping abruptly. In federated learning, each client often trains only on short, recent fragments of a user's history, and many users interact infrequently, which makes the model overfit short-term fluctuations and ignore long-term patterns. To alleviate this, we add a temporal consistency regularizer that encourages smooth changes in user embeddings across adjacent subsequences of the same user.

\begin{figure}[!h]
\centering
\includegraphics[width=0.9\linewidth]{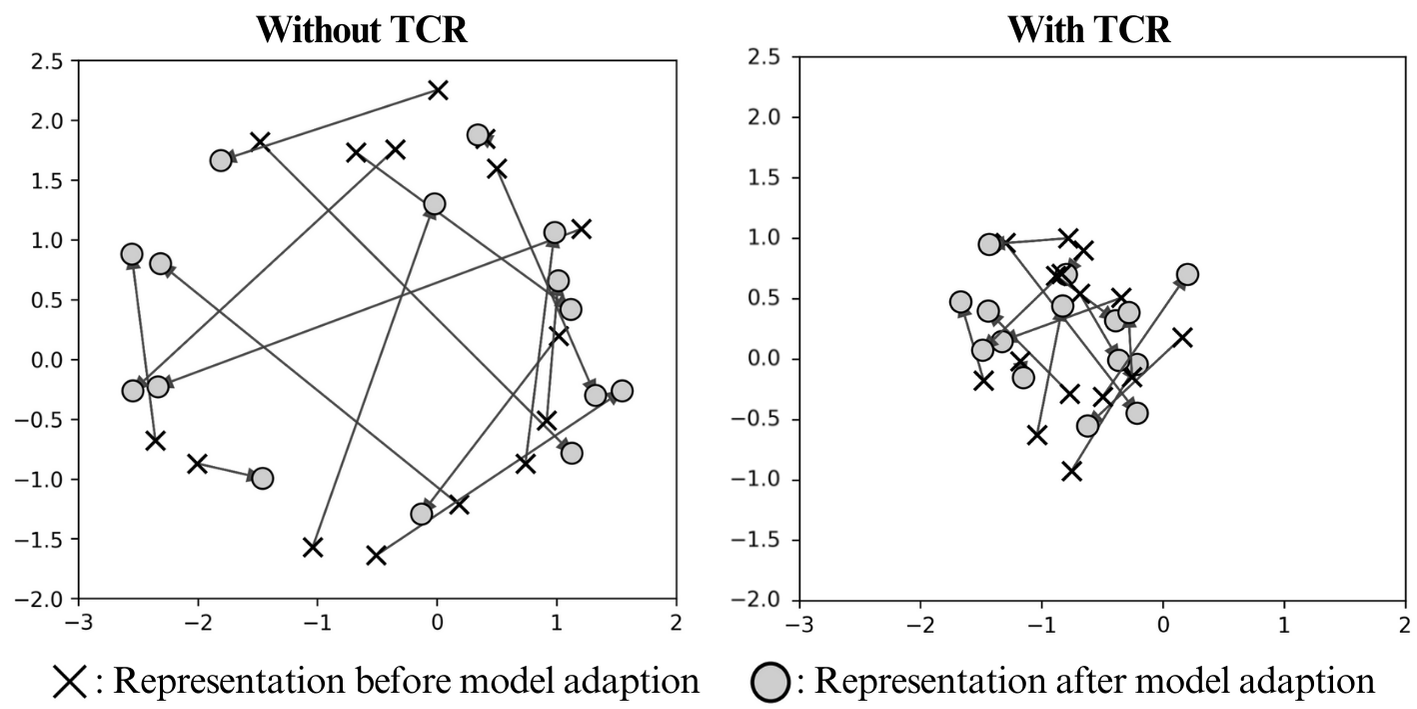}
\caption{Impact of temporal regularization on users' sequence embeddings.}
\label{fig:impact_tcr_embedd}
\end{figure}

For a user $u$ with interaction sequence $S_u$, we extract two temporally adjacent subsequences $S_u^{(t)}$ and $S_u^{(t+1)}$ and define:
\[
\mathcal{L}^{\text{TCR}}_u = \left\| f_\theta(S_u^{(t)}) - f_\theta(S_u^{(t+1)}) \right\|_2^2.
\]
This loss penalizes abrupt shifts in the latent space so that short-term noise does not dominate long-term preferences. As shown in \autoref{fig:impact_tcr_embedd}, temporal regularization produces smoother, more coherent user-embedding trajectories.

\subsection{Client-Side Optimization Objective}

Combining the above components, each client optimizes a single objective that includes next-item prediction, multi-view contrastive alignment and temporal consistency. For each user $u$ we minimize:
\[
\mathcal{L}_{u} = \mathcal{L}^{\text{Rec}}_{u} + \lambda_{\text{CL}} \cdot \mathcal{L}^{\text{CL}}_{u} + \lambda_{\text{TCR}} \cdot \mathcal{L}^{\text{TCR}}_{u}.
\]
Here, $\mathcal{L}^{\text{Rec}}_{u}$ is the next-item prediction loss, $\mathcal{L}^{\text{CL}}_{u}$ aggregates the sequence, user and item contrastive terms, and $\mathcal{L}^{\text{TCR}}_{u}$ is the temporal regularizer. The hyperparameters $\lambda_{\text{CL}}$ and $\lambda_{\text{TCR}}$ control the strength of contrastive alignment and temporal smoothing, so that the local model can balance accuracy, robustness to noise and stability of preference trajectories before aggregation.
\todo{
Overall, the contrastive learning strategy in \toolname{} differs from existing federated approaches by combining local multi-view augmentation, sequential modeling, and temporal regularization, rather than relying on local--global alignment or server-side denoising alone.
}

\subsection{Adversarial Robustness and Server-Side Defense}
\label{server_regularizer}

Local regularization improves robustness on individual clients, but federated recommendation remains exposed to cross-client poisoning attacks that act through aggregation. As illustrated in \autoref{fig:framework}, before contrastive learning the popular items are concentrated in a compact cluster in the embedding space, making it harder for attackers to reach this region. After contrastive learning, their embeddings become more dispersed and closer to a roughly normal distribution, so poisoned items can more easily appear near popular clusters and gain exposure in the ranking. To counter this, we design a server-side defense with two terms: a contrastive separation loss that pushes suspicious items away from popular ones, and a variance regularizer that keeps popular clusters tight. Promotion attacks inject poisoned interactions so that the embeddings of target items move toward those of popular items and increase their rank. We discourage this by enforcing a margin between globally popular items $\mathcal{V}_{\text{hot}}$ and suspicious or low-visibility items $\mathcal{V}_{\text{sp}}$ with a separation loss:
\[
\mathcal{L}^{\text{sep}} = \sum_{v_i \in \mathcal{V}_{\text{hot}}} -\log \frac{\exp(1/\tau)}{\sum_{v_j \in \mathcal{V}_{\text{sp}}} \exp(\text{sim}(v_i, v_j)/\tau)}.
\]
This makes it harder for malicious items to mimic the representations of truly popular items.

Camouflage attacks try to embed bad items inside high-visibility clusters so that they appear in ``similar items'' lists without obvious rank spikes. To reduce this effect, we minimize the intra-cluster variance of popular items:
\[
\mathcal{L}^{\text{var}} = \sum_{v_i \in \mathcal{V}_{\text{hot}}} \operatorname{Var}\left( \{ f_\theta(v_j) \mid v_j \in \mathcal{N}(v_i) \} \right),
\]
where $\mathcal{N}(v_i)$ is the neighborhood of items near $v_i$ in embedding space. Keeping variance low discourages anomalous items from blending into these clusters.

The server-side objective combines both defenses:
\[
\mathcal{L}^{\text{server}} = \lambda_{\text{sep}} \cdot \mathcal{L}^{\text{sep}} + \lambda_{\text{var}} \cdot \mathcal{L}^{\text{var}}.
\]
Here, $\lambda_{\text{sep}}$ and $\lambda_{\text{var}}$ weight the separation and variance terms. Together with the client-side item view, this provides coordinated local and global protection against promotion and camouflage attacks.

\begin{algorithm}[!ht]
\caption{Federated Training with \toolname}
\label{alg:fortress}
\begin{algorithmic}[1]
\REQUIRE Total communication rounds $T$, local epochs $E$, learning rate $\eta$, hyperparameters $\lambda_{\text{CL}}, \lambda_{\text{TCR}}, \lambda_{\text{sep}}, \lambda_{\text{var}}$
\STATE Initialize global model parameters $\theta^{(0)}$
\FOR{$t = 1$ to $T$}
    \STATE Server samples a subset of users $\mathcal{U}_t$
    \FORALL{user $u \in \mathcal{U}_t$ \textbf{in parallel}}
        \STATE Client creates contrastive augmentations:
        \STATE \hspace{1em} sequence views $\tilde{S}_u^{(1)}, \tilde{S}_u^{(2)}$,
        user views $\mathbf{u}_i', \mathbf{u}_i''$,
        item views $\mathbf{v}_k', \mathbf{v}_k''$
        \STATE Client extracts temporally adjacent subsequences $S_u^{(t)}, S_u^{(t+1)}$
        \FOR{$e = 1$ to $E$}
            \STATE Update $\theta_u$ by minimizing:
            \\ \hspace{1em}
            $
            \mathcal{L}_u = \mathcal{L}^{\text{Rec}}_u + \lambda_{\text{CL}} \mathcal{L}^{\text{CL}}_u + \lambda_{\text{TCR}} \mathcal{L}^{\text{TCR}}_u
            $
        \ENDFOR
        \STATE Client uploads $\theta_u$ to server
    \ENDFOR
    \STATE Server aggregates uploaded models: 
    \\ \hspace{1em}
    $
    \theta^{(t+1)} \leftarrow \text{Aggregate}(\{\theta_u\}_{u \in \mathcal{U}_t})
    $
    \STATE Server computes:
    \\ \hspace{1em}
    $
    \mathcal{L}^{\text{server}} = \lambda_{\text{sep}} \mathcal{L}^{\text{sep}} + \lambda_{\text{var}} \mathcal{L}^{\text{var}}
   	$
    and updates item embeddings by gradient descent on $\mathcal{L}^{\text{server}}$.
\ENDFOR
\end{algorithmic}
\end{algorithm}

\section{Conclusion}
\label{sec:conclusion}

We propose \toolname, a federated sequential recommender that jointly tackles data sparsity and model poisoning attacks. \toolname integrates multi-view contrastive learning on clients, a temporal consistency regulariser to stabilise user representations, and a popularity-aware server-side defence that constrains suspicious promotion patterns in the global embedding space. 
Future work will extend \toolname to more expressive backbones, richer hybrid attack models, and real-world cross-device deployments with tighter communication and availability constraints~\cite{nguyen2026handling,pham2025multilingual,pham2025extensible,nguyen2025device,nguyen2026review,nguyen2024manipulating,nguyen2025privacy}.

\end{document}